# Mechanism of Organization Increase in Complex Systems

Georgi Yordanov Georgiev[1,2,3*], Kaitlin Henry[1], Timothy Bates[1], Erin Gombos[1,4], Alexander Casey[1,5], Michael Daly[1,6], Amrit Vinod[1,7] and Hyunseung Lee[1]

[1]Physics Department, Assumption College, 500 Salisbury St, Worcester, MA, 01609, USA
[2]Physics Department, Tufts University, 4 Colby St, Medford, MA, 02155, USA
[3]Department of Physics, Worcester Polytechnic Institute, Worcester, MA, 01609, USA
[4]Current address: National Cancer Institute, NIH, 10 Center Drive, Bethesda, MD 20814, USA
[5]Current address: University of Notre Dame, Notre Dame, IN 46556, USA
[6]Current address: Meditech, 550 Cochituate Rd, Framingham, MA 01701, USA
[7]Current address: University of Massachusetts Medical School, 55 Lake Avenue North, Worcester, MA 01655, USA

[*]Corresponding author. Emails: ggeorgie@assumption.edu; georgi@alumni.tufts.edu;

**Abstract**

This paper proposes a variational approach to describe the evolution of organization of complex systems from first principles, as increased efficiency of physical action. Most simply stated, physical action is the product of the energy and time necessary for motion. When complex systems are modeled as flow networks, this efficiency is defined as a decrease of action for one element to cross between two nodes, or endpoints of motion - a principle of least unit action. We find a connection with another principle that of most total action, or a tendency for increase of the total action of a system. This increase provides more energy and time for minimization of the constraints to motion in order to decrease unit action, and therefore to increase organization. Also, with the decrease of unit action in a system, its capacity for total amount of action increases. We present a model of positive feedback between action efficiency and the total amount of action in a complex system, based on a system of ordinary differential equations, which leads to an exponential growth with time of each and a power law relation between the two. We present an agreement of our model with data for core processing units of computers. This approach can help to describe, measure, manage, design and predict future behavior of complex systems to achieve the highest rates of self-organization and robustness.

## 1. Introduction

We study the processes of self-organization in nature seeking to find one set of rules or one universal law that causes all of them to occur. The importance of such an endeavor has been recognized, particularly when it comes to the optimization of energy flows in a system (Hubler and Crutchfield, 2010, Chaisson, 2004, 2011a.b). The appearance of elementary particles from radiation energy after the Big Bang started a chain of events that is the object of our study. Those particles assembled into atoms, which in turn formed molecules that gave rise to organisms, and eventually our human civilization and beyond. We see this chain of hierarchical events unified by the same underlying natural laws, leading to the rise of all of them. The endeavor to find a unifying theory for self-organization is an exciting prospect that we hope will continue motivate others to continue in this endeavor.

There are many crucial questions that urgently need an answer. What principle determines the motions in complex systems? Which motions are preferred? What does it mean for a system to be organized and to self-organize? Why do systems self-organize? How do we measure organization? What does self-organization depend on? What is the rate of self-organization and when does it stop or reverse? How does the increase of the size of a system affect the increase of its organization? Why do some complex systems continue to self-organize for billions of years but others are temporary? What is special about systems that reach the highest levels of organization? What is driving them forward in their increased levels of organization? These and other similar questions have been staring at us since we became conscious, but we do not yet know all of the answers. We address some of them in this paper.

Our approach is to study the minimization of physical action per unit motion and maximization of the total action for all motions in systems. Self-organization in complex systems can be described as an increased efficiency of physical action, which provides a means to define what exactly organization is and how it is achieved and measured (Georgiev and Georgiev, 2002; Georgiev, 2012; Georgiev et al., 2012). This approach stems from the principle of least action, which underlies all branches of physics and all motion in nature. Complex systems are comprised of individual elements. Each element is the smallest mobile unit in the system and moves, most often, in a flow of other elements along a network of paths (edges) between the starting and ending points in order to build, recombine, or change the system. In CPU's, one unit of motion (event) is a single computation in which electrons flow from the start node to the end node. In our model, the flow is based off of events, not of energy or matter per se, even though they are participating in the events. It has been shown that in complex systems the nodes of its network representation need to be well defined, so the elements can traverse deterministic walks instead of random walks (Boyer and Larralde 2005). Random walks characterize equilibrated, non-self-organizing system.

In complex systems, elements cannot move along their least possible action paths that characterize their motion outside of systems because of obstacles to the motion (constraints). The principle of least action expanded for complex systems states that systems are attracted toward a state with least average action per one motion given those constraints (Georgiev and Georgiev, 2002; Georgiev, 2012; Georgiev et al., 2012). Similarly, the Hertz's principle states that objects move along paths with the least curvature (Hertz, 1896; Goldstein, 1980), and the Gauss principle states that they move along the paths of least constraint (Gauss, 1829). We extend these two principles for complex systems that the elements do work on the constraints to minimize them, reducing the curvature and the amount of action spent for unit motion. The new geodesic in the curved by the constraints to motion space is the path with the lowest amount of action. When the elements do work to minimize the constraints, they form paths of least constraint. Those are the flow paths in the system (Georgiev and Georgiev, 2002; Georgiev, 2012; Georgiev et al., 2012). Because the action is lowest along those paths of least action, compared to all neighboring paths, the rest of the elements traversing the same nodes, move along the same paths. As the constraints are minimized further, and the action decreases along a certain path, the probability for more elements to move along them increases, and those paths become attractors for other elements, which further minimizes action along them. Indeed, it has been recognized that in complex systems the major control parameter is the throughput (Hübler, 2005). Therefore in our work, organization is defined as the state of the constraints to motion determining the average action per one element of the system and one of its motions. As the constraints are minimized, the same motion is done more efficiently, i.e. the same two nodes are connected using less action, and organization increases.

We use a flow network representation of a complex system, where the trajectories of the elements are along flow paths of least action, compared to neighboring paths. A flow network implies an inflow and outflow of energy and can exist only in open systems far from equilibrium. The sources and sinks define the start and endpoints of the elements and flows in the system, which are the nodes of the network. As a result of the "principle of minimum dissipation per channel length" natural network formation "in an open, dissipative system" is as "branching, hierarchical networks" (Smyth and Hubler, 2003) which means that flow networks maximize their energy efficiency if they have a hierarchical, fractal structure. This points to an explanation from first principles of the formation of hierarchical fractal flow networks, such as internet, transportation networks, respiratory or cardiovascular systems, and many others, which "share the scale-free property … of self-similarity or fractality" (Rozenfeld-11). We see evidence of fractality in scale free systems in nature everywhere, from snowflakes and coast lines to data and molecules, and recent research has begun to quantify these observations (Rozenfeld-11). The network model of a complex system has gained importance in recent years (Alain et al., 2008; Ángeles et al., 2007; Dangalchev, 1996; Liu et al., 2013; Mark et al., 2011; Wu et al., 2006; Xulvi-Brunet and Sokolov, 2007). The scaling laws of transport properties for scale free networks have been found (Goh et al., 2001) and their betweenness centrality was measured (Kitsak-07). The self-similar scaling of density was found important in complex real-world networks (Blagus-12). It was found that scale-free networks allow routing schemes to self-adjust in order to overcome congestion (Zhang-07, Tang-09). Congestion is a jamming transition which decreases the flow, which in turn lowers the action efficiency of the system, therefore its level of organization. This leaves us with questions about why these fractal, scale-free flow networks exist in a first place, as opposed to elements moving by diffusion or in a different network pattern? When the system's size is below a certain threshold, diffusion must minimize the action for the motions in the system given their constraints. When those systems are above certain size, it should take less action to move along flow channels where the constraints to motion are minimized, as compared to diffusion. This hints about a size dependence of complexity and efficiency of flow networks. As networks grow, apparently the structure that is most efficient in terms of unit action and most ubiquitous is the scale free.

Size-Complexity relation: Self-organizing systems have two attractors – increase of level of complexity and increase in size of the system (Bell and Mooers, 1997; Bonner, 2004; Carneiro, 1967). Represented as flow networks, we measure those two quantities as a decrease in average unit physical action (the average action necessary for one element of the system to cross one edge) and an increase in total physical action (the sum of the actions of all elements in the system in certain interval of time). The positive feedback between the least unit action and the maximum total action leads to a process of exponential growth in time of both of them and a power law relation between them characterizing developing systems and is ubiquitous in nature (Bell and Mooers, 1997; Bertalanffy, 1968; Bonner, 2004; Carneiro, 1967). The effect of system's size on its efficiency also supports this observation (Bejan 2011, Kleiber-1932, West 1999) as do the processes of growth and efficiency in information and other technologies (Moore 1965, Kurzweil 2005, Nagy 2011). Development and evolution in different systems have been studied through the dynamics of complex systems in great detail, and some projections about the future of this trajectory have been made (Bar-Yam, 1997; Gershenson and Heylighen, 2003; James et al., 2011; Salthe, 1993; Smart, 2002; Vidal, 2010). In describing the emergence of order in random systems, simulations have proved particularly useful (Kassebaum and Iannacchione, 2009) which will be the next step in this research, after the mechanisms of self-organization are understood sufficiently.

In this paper we strive to show that the quality of a system depends on its quantity and when one is increased or decreased the other is affected in the same way i.e. they are in a positive feedback loop. We also want to show that this dependence is a major driving force and a mechanism of progressive development measured as increase in action efficiency in complex systems. This is important for understanding self-organization and can be used in designing more efficient complex systems. We hope that researchers in various disciplines from chemistry and biology to engineering and social science will be able to apply the results.

## 2. Theory

### 2.1 Overview

This manuscript proposes a variational approach, maximizing total action and minimizing the average unit action with increase of level of organization in complex systems. The total quantity of action is the sum of all quanta of action occurring in a system per unit time. Quanta of action are obtained as the total action is divided by the Planck's constant – the smallest unit (quantum) of action in nature. The average unit action is the average action necessary for one event in a system, measured as the total action is divided by the total number or events in certain interval of time. Lower average unit action means higher action efficiency of a system, which is our definition for quality and organization, as inversely proportional to the average number of quanta of action per unit motion. The importance of variational (extremization) principles in describing systems' behavior, and the extension of the principle of least action for dissipative systems have been noted widely (Sieniutyez and Farkas, 2004). Eric Chaisson's work points to maximization of the free energy rate density (FERD), i.e. the flow of energy per unit mass and time in systems, as a function of time (Chaisson, 1998, 2001). He describes how the evolution of systems of different nature, such as physical, chemical, biological and social, correlates with an exponential increase of the FERD, which indicates that as systems are becoming more organized they are moving further away from thermodynamic equilibrium. This supports the observation of increase of quantity in systems, as increased energy gradients and therefore flows, through more organized systems. There are many other variational principles which have been applied to self-organization, such as the maximum entropy production (Paltridge 1979) the minimum entropy production (Nicolis and Prigogine, 1977) and the least dissipation principles (Onsager and Machlup, 1953). The minimization of constraints to motion of flows of elements is also supported by the constructal law which is that complex systems' configurations evolve in a way that provides easier access to the currents that flows through them (Bejan, 2005) which is also an extremization principle. The utility of the principle of least action to describe self-organization has been shown (Annila and Salthe, 2010; Annila, 2010; Mäkelä, and Annila, 2010; Chatterjee, 2012, 2013) and has been used to describe a "natural selection for least action" (Pernu and Annila 2012, Hartonen and Annila 2012). Hubler and Crutchfield noted a "tendency in systems with a constant flow to minimize energy consumption" (Hubler and Crutchfield, 2010) and introduced a "principle of minimum dissipation per channel length" (Smyth and Hübler, 2003). This is not far from our view of minimization of time use and energy dissipation per unit motion connected to a maximization of total energy dissipation and time of existence of the entire system. This leads to a minimum entropy production per unit motion, related to a maximum total entropy production of the system. A minimization of an economy function has also been used (Boyer and Larralde 2005). It has been suggested that applying the principle of least action to biological systems could be very helpful in solving the mystery of how, during organism development and differentiation, complex patterns emerge (Vandenberg et al., 2012).

To clarify why we minimize the action but not just the energy or time per unit motion, let's suppose that we minimized them separately. If only free energy is minimized, the result is that motion ceases and flows stop. This is the process of crystallization as one example. Therefore if free energy was minimized, it would take an infinite amount of time for an element to bridge two nodes, i.e. complex systems cannot exist and function. There is no flow of energy through them. Similarly, minimizing only time would lead to a paradox, necessitating infinite amounts of energy to be spent by the system as the time interval for an event approaches zero and the speed of the elements increases to the speed of light. Time and energy must therefore be in balance with each other, which is achieved by the principle of least action or the least product of time and energy, but not the least amount of each of them separately.

To specify the systems studied, we need to separate self-organizing complex systems in two classes: passive and active. Passive are those that exist until external energy gradients (differences) are equilibrated. Those energy gradients occur independently of the system. Temporary dissipative structures, such as Bernard cell or vortices belong to this type. They minimize unit action to accommodate flows while energy differences exist, but do not exhibit further self-organization, and fall apart as the energy gradients are equilibrated. Active systems are those that increase their energy gradients and drive themselves further out of equilibrium, actively increasing their size and action efficiency further - evolving in time. This active type of systems exhibits continuous self-organization, growth and increased robustness, as in biological and social systems (Bar-Yam, 1997; Bertalanffy, 1968; Chaisson, 2001, Gershenson and Heylighen 2003).

Outside of complex systems, the physical action has a single minimum for the motion of a particle compared to all other paths, given by its equations of motion, which is along a geodesic. In complex systems, due to the constraints to motion, moving along the same geodesic will have higher action, compared to a large set of symmetric longer trajectories. For infinitely long paths, action rises to infinity. In phase space, this forms a well-known "Mexican hat" surface with a circular minimum around the geodesic of a particle which describes its motion when it is not a part of a system. On this surface the elements of the system spontaneously choose one of the infinite number of minimum action trajectories, which signifies a phase transition from a simple to a complex system, or from one level of organization to another. The initial geodesic paths of the elements become the "vacuum" or "ground state" of the complex system. There is a body of work on phase transitions and symmetry breaking in open dissipative systems with local interactions and global constraints, causing bottom-up and top-down sequences of symmetry breakings (Hübler, 2005). Lower symmetry is generally connected to more order, and higher symmetry, to more randomness and entropy. An order parameter approach for describing these symmetries has been proposed to measure self-organization (Haken 1982, 2006).

## 2.2 Basic Principles

The principle of least action determines the motions of all objects in the universe, therefore it must determine the motions of elements in complex systems. We explore the idea that self-organization is driven by the principle of least action for systems, which states that the variation of the average unit action is zero in the most organized state,

$$\delta \frac{\sum_{ij}^{nm} I_{ij}}{nm} = 0$$

where $\sum_{ij} I_{ij}$ is the total amount of action in the system per unit time. n is the total number of elements in the system, and m is the number of edge crossings of one element per unit time.

When the unit action decreases, the system becomes more efficient, obeying the principle of least action, i.e. self-organizes:

$$\delta \frac{\sum_{ij}^{nm} I_{ij}}{nm} < 0$$

As this is a minimization problem, the limit is a minimum, which approaches zero when the system is infinitely organized:

$$\lim_{t \to \infty} \frac{\sum_{ij}^{nm} I_{ij}}{nm} = \min$$

When the unit action increases, its level of organization decreases:

$$\delta \frac{\sum_{ij}^{nm} I_{ij}}{nm} > 0$$

When a system is regressing, it can reach a limit which approaches infinity in the maximally random entropic state:

$$\lim_{t \to \infty} \frac{\sum_{ij}^{nm} I_{ij}}{nm} = \max$$

The total action in a system is the sum of all actions of all elements and edges that they are crossing. i.e. it is the action necessary for the total flow in the system as measured by all events that are occurring along the flow network. Quality and quantity are proportional in our model therefore when the total action increases, the system self-organizes, as inferred from the literature and our data:

$$\delta \sum_{ij} I_{ij} > 0$$

When the total action decreases, the organization decreases:

$$\delta \sum_{ij} I_{ij} < 0$$

Self-organization is a maximization problem, therefore the limit is a maximum during self-organization which approaches infinity when the system is infinitely organized.

$$\lim_{t \to \infty} \sum_{ij}^{nm} I_{ij} = \max$$

In that state, the action cannot increase futher and its variation is zero:

$$\delta \sum_{ij}^{nm} I_{ij} = 0$$

This is the final state, where a system cannot organize anymore, i.e. cannot grow in quality or quantity.

The reverse is also true: when a system becomes less organized, it falls apart and decreases in size, ultimately until no elements are left in it, and it is not defined as a system anymore, i.e. its total action is decreasing to a minimum, which is zero when the system stops to exist.

This leads us to a measure for organization which is inversely proportional to the average number of quanta for an element in a flow network to cross one edge – its action efficiency. For identical elements (Georgiev, 2012):

$$\alpha = \frac{hnm}{\Sigma_{ij} I_{ij}}.$$

Where h is the smallest unit of action in nature, the Planck's constant.

We recognize that nm is the total number of edge crossings, the flow $\phi$ of elements per unit time in the flow network, $\phi$=nm. $Q = \frac{\Sigma_{ij} I_{ij}}{h}$ is the total number of quanta in the system in certain interval of time. Therefore we can rewrite the expression for α in the following way:

$$\alpha = \frac{\phi}{Q}.$$

A lower limit: when the system is at equilibrium, the entropy is at a maximum and therefore the organization is zero. The random motion of the elements determines paths that are non-directed and become infinitely long if they have to cross between the same two points as defined in a flow network. This makes the action per crossing infinity, and the organization zero.

$$\alpha = \frac{hnm}{\sum_{i,j} I_{i,j}} = \frac{hnm}{\infty} = 0$$

This limit is also zero for finite total action in the system, when there is no flow in the system, i.e. there are no elements, n=0, or the elements do not have energy to cross any edges, m=0.

An upper limit: The organization becomes unity when one quantum of action per event is used. One quantum of action is the smallest possible unit of action.

$$\alpha = \frac{hnm}{\sum_{i,j} I_{i,j}} = \frac{hnm}{hnm} = 1$$

Possible exceeding the upper limit: In parallel processes, such as quantum computing, when with one quantum of action an electron traverses many edges, doing a computation along each one, or alternatively if elements cross several nodes in the network with one quantum of action, the ratio becomes very large and the process of organization can continue to infinity.

$$\lim_{t \to \infty} \alpha = \infty$$

i.e. the final states of variation, when the system cannot organize anymore, which is:

$$\delta \frac{\sum_{ij}^{nm} I_{ij}}{nm} = \delta \sum_{ij}^{nm} I_{ij} = 0$$

will never be reached.

### 2.3 A positive feedback loop

We explore the model that to achieve higher levels of action efficiency per unit motion (quality), the system needs to utilize more total action, i.e. more energy and time (quantity). An increase in $Q$ of a system, drives it further away from equilibrium, allowing more work to be done to minimize the constraints, leading to less unit action and higher levels of organization. At the same time, in order for the system to grow in size, spread, accumulate more matter and energy, and be able to sustain higher energy flows through it, it needs to be at higher levels of organization, or more efficient in its unit action. Otherwise the increased quantity of flow through the system will destabilize it. The reasons can be many: increase of energy dissipation, jamming, and others. One of the effects of increased action efficiency is that energy dissipation for unit motion is decreased, as it is later evident in the data for CPUs. This implies a positive feedback between the quality and quantity of a self-organizing system.

Let's investigate this model to study how $\alpha$ and $Q$ depend on each other and what time behaviour can be predicted. We explore a positive feedback loop between the quality and quantity.

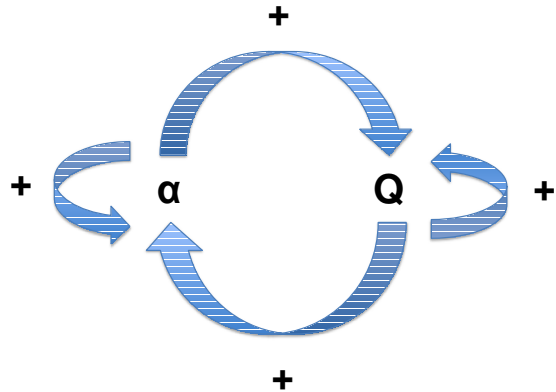

Figure 1: A positive feedback loop between $\alpha$ and $Q$. The possible positive feedbacks are between $\alpha$ and $Q$, and $Q$ and $\alpha$, where they induce growth in each other, and of $\alpha$ and $Q$ on themselves, as shown. Each of these loops affects the growth of $\alpha$ and $Q$, which is expressed by their time derivatives.

### 2.4. Differential Equations and Solutions for $\alpha$ and $Q$

Let's set $\dot{\alpha}$ and $\dot{Q}$ as the time derivatives of $\alpha$ and $Q$. The positive feedback produces a system of simultaneous differential equations, if we include the cross-terms and self-terms between the two quantities:

$$\dot{\alpha}=a_{11}\alpha+a_{12}Q$$
$$\dot{Q}=a_{21}\alpha+a_{22}Q$$

Where $a_{ij}$ are coefficients. To prevent self-growth, which will mean that systems can grow in size without increasing in organization, or that they can increase in their levels of organization without increasing in size, both of which are not observed in nature, we set the self-terms to zero by $a_{i=j} = 0$, and the system becomes:

$$\dot{\alpha} = a_{12} Q$$
$$\dot{Q} = a_{21} \alpha$$

The numerical solution of this system produces a strictly exponential curve, shown below.

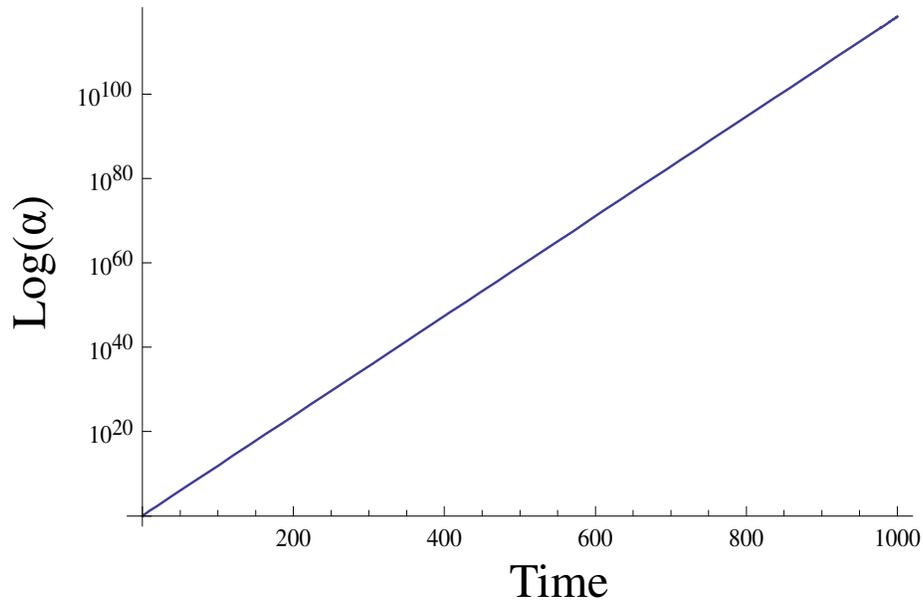

**Figure 1.** Numerical solution of organization vs. time from the above system of differential equations, in arbitrary units. The result is identical if quantity (Q) vs. time is plotted.

The analytical solution can be simplified for t>1, to:

$$\alpha = \alpha_0 e^{\tau t}$$
$$Q = Q_0 e^{\beta t}$$

This exponential growth in time is verified by the data for CPUs in Figs. (2) and (3). Both quantities are fit with an exponential, on log-linear plots.

Eliminating time, a solution of the exponential equations is the power law relation:

$$\alpha = \eta Q^{\gamma}.$$

Where $\gamma = \frac{\tau}{\beta}$. and $\eta = \frac{\alpha_0}{Q_0^{\gamma}}$.

This is evident on Fig. (4), where organization increases as a power law, as a function of the quantity of the system.

Dividing the differential equations for the growth of organization and quantity produces:

$$\dot{\alpha} = \frac{\tau}{\beta}\frac{Q}{\alpha}\dot{Q}.$$

The meaning of this result is that if the total number of quanta in the systems (the total quantity) is fixed, then the quality (the level of organization) cannot change. In other words, the system cannot become better organized if the total amount of action in the system, or its quantity is fixed. The previous two equations carry the information that if $Q$ is decreased or increased, the level of organization $\alpha$ will decrease or increase as well, and vice versa. This confirms the observations about the size-complexity (Bonner, 2004; Carneiro, 1967; Bell and Mooers, 1997) and and size-efficiency rules (Bejan 2011, Kleiber-1932, West 1999). For continuing increase of the level of organization, increase of the size of the system is necessary, and vice versa.

3. **Data and Methods**

Data were collected from Intel Corporation Datasheets (Intel Corporation). The Instructions Per Second (IPS) for each processor was divided by the Thermal Design Power (TDP) as a measure of the total power consumption by the CPUs at maximum computational speed, for consistency. The result was multiplied by the table value of the Planck's constant, $h = 6.626\ 10^{-34}\ (Js)$, as the smallest quantum of action, to solve for $\alpha$, as the inverse of the number of quanta of action per one instruction per second. The TDP was divided by the $h$ value to find the total number of quanta of action per second (Q). Only processors for desktops or laptops were used, because some of the specialized processors, such as for phones or tablets, perform slower in order to use less energy and fall below the trend line.

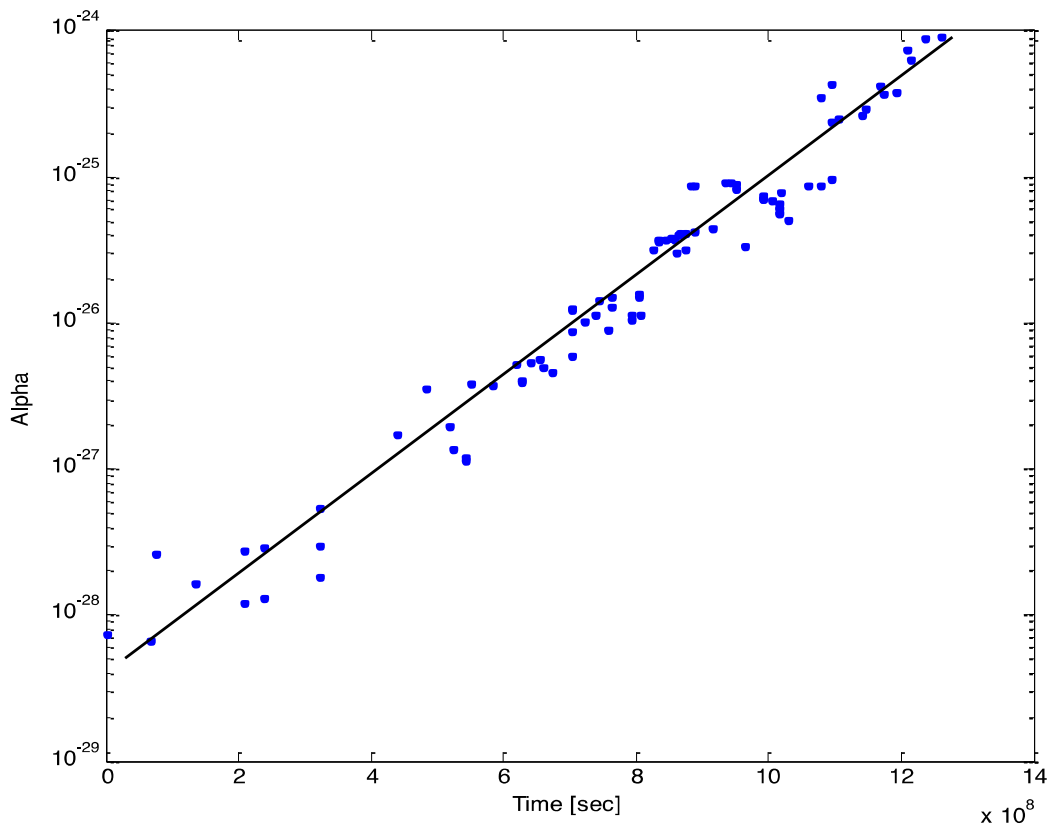

**Figure 2.** Data for α versus time for CPUs from 1971 to 2011 (closed circles) and an exponential fit (solid line). The transition from single to multicore processors around time $10^9$[sec], does not affect the trend. α does not increase smoothly but in steps.

4. **Results:**

As shown in Fig. 2, α increases exponentially by more than four orders of magnitude since the first microprocessors were introduced, which matches well with the theoretical prediction for exponential growth. Some of the possible reasons for the action efficiency increase per computation, as reflected in this trend for increased organization are: the nodes are closer together due to the miniaturization of the transistors on the CPUs; the elements are smaller (smaller packages of electrons); the constraints are smaller (less friction and therefore energy dissipation per one computation); the curvature of the paths of the electrons for one computation is smaller; and others. All these reasons lead to the effect that less time and energy is used for crossing the edges between the nodes necessary for computations. At the same time the total number of quanta of action in the system increases exponentially by more than two orders of magnitude, as can be seen on Fig. 3, which also matches well with the theoretical prediction for exponential growth. Another good correspondence with the model is that the relation between α and Q is a strict power law, indicating a positive feedback between the two and confirmed by the data on Fig. 4. On all three graphs there are also visible oscillations around the best fit values. The values for all constants in the equations were determined from the lines of best fit and are presented in Table 1.

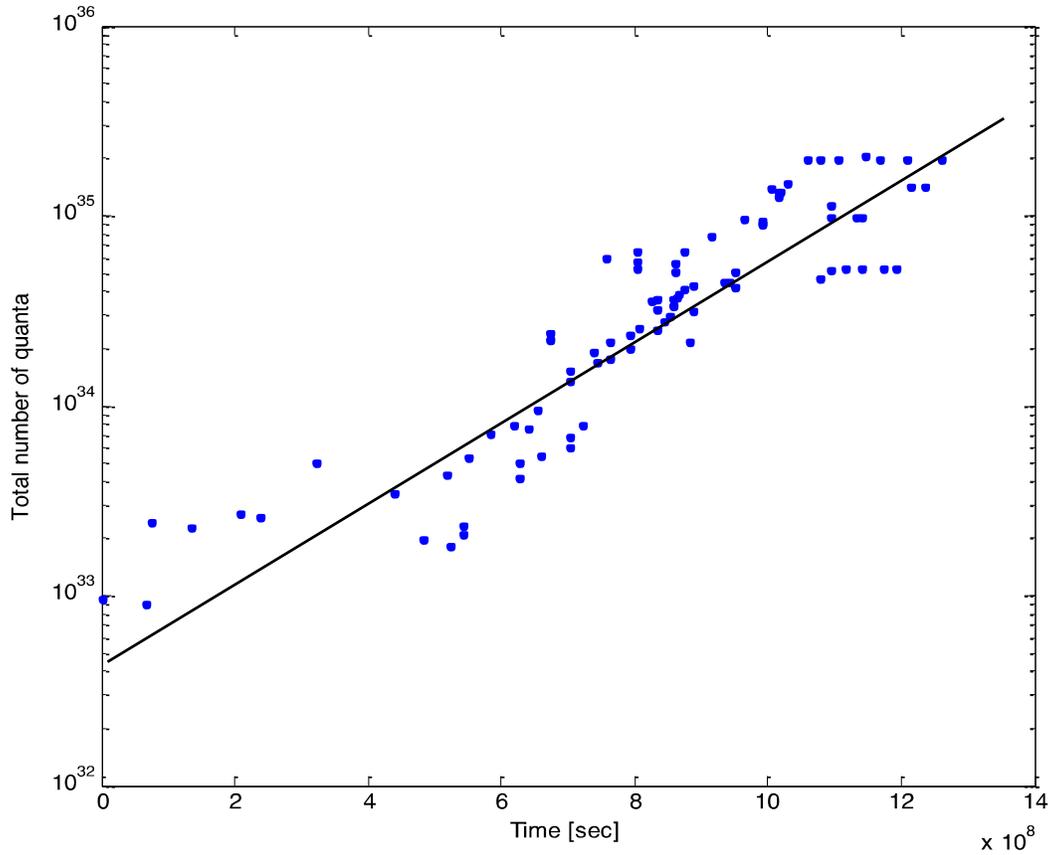

**Figure 3.** Data for Q versus time for CPUs from 1971 to 2011 (closed circles) and an exponential fit (solid line). Note variations around the average growth rate.

Table 1. Values of constants.

| Constant | $\tau$ | $\beta$ | $\gamma$ | $\alpha_0$ | $Q_0$ | $\eta$ |
|---|---|---|---|---|---|---|
| Value | $6.5 \times 10^{-9}$ | $3.9 \times 10^{-9}$ | 1.3775 | $4 \times 10^{-29}$ | $7 \times 10^{32}$ | $6 \times 10^{-74}$ |

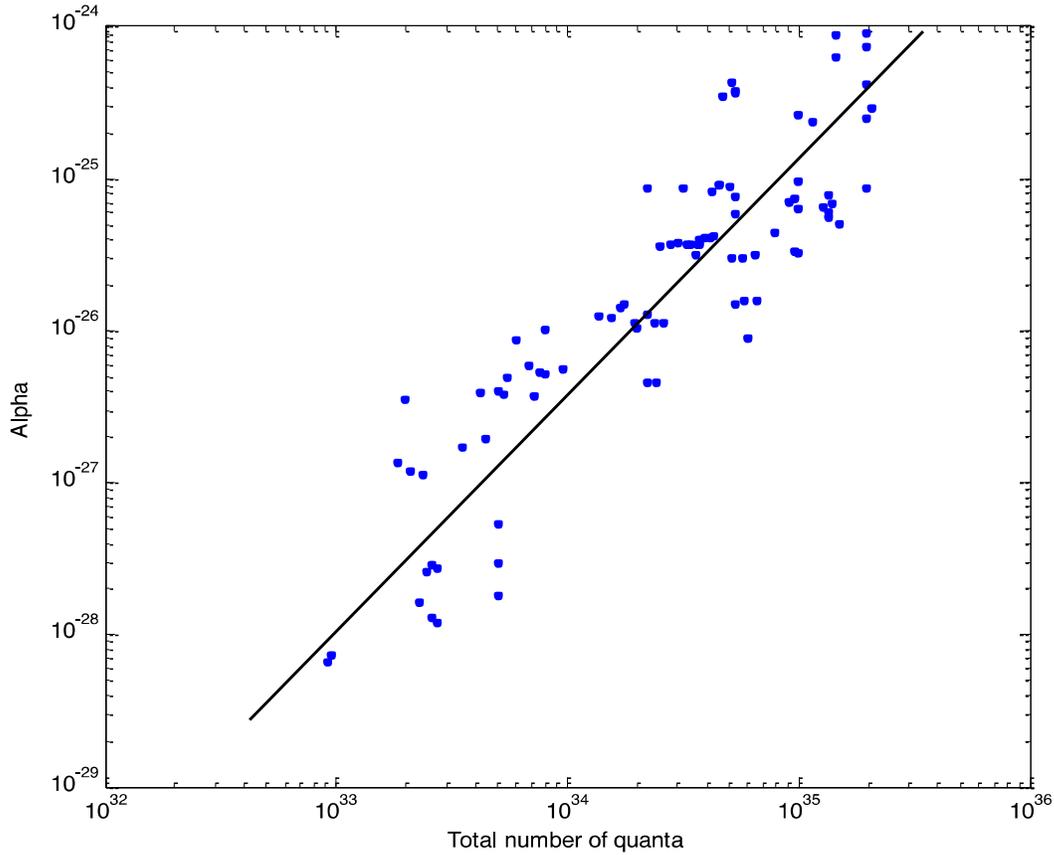

**Figure 4.** Data for α and Q from 1971 to 2011 (closed circles) and a power law fit (solid line). Note variations around the average.

5. **Summary and Discussion**

In this paper, the broadened principle of least action was applied to a sample complex system of CPUs, and the decrease of unit action was studied as function of total action. The state of the system with lower average unit action was defined to be better organized which provides a numerical measure for organization. Level of organization and quantity were found to be in a positive feedback, and to increase exponentially in time, and as a power law of each other, with observed periodic variations around the averages. We defined two measures, α and Q, for the process of growth and organization of complex flow networks, which represent general self-organizing complex systems. This types of flow networks were modeled with a system of ordinary differential equations for positive feedback between α and Q of the system. We obtained exponential growth solutions for α and Q as a function of time, and a power law relation between the two. We fit the data for organization α (quality) versus total number of quanta of action in the system Q (quantity) for CPUs, which confirmed the prediction. The system of equations also predicts that due to the positive feedback nature of the interaction of quality and quantity, if one of the functions α or Q decrease, that will lead to the decrease of the other; if one of them increases the other increases; and if one of them is fixed, the other will reach a constant value.

We want to clarify that based on the power law relation between α and Q we do not imply that one of them can be discontinuously increased or decreased by an arbitrary amount, for the data studied, which will automatically bring a proportional change in the other. On the contrary, increase in organization is a time dependent process, and the changes in α and Q must occur within small deviations of their power law relation. This is evident in the oscillations of both α and Q around their power law relation on Figure 4 visible as periodic fluctuations around the average fit. These oscillations in the growth of α and Q are also visible in their time dependencies Figure 2 and Figure 3. The data vary around the average fits from the model, which lead us to an explanation of the underlying mechanism of quality-quantity oscillations around an average rate of increase of quality as a function of quantity. The increase of one allows and stimulates the increase of the other, within certain limits. We interpret this in the following way: adding new elements to the system allows and forces the increase of its organization. One way that this can happen is by grouping of the elements allowing faster and more energy efficient constraint minimization per element (Georgiev, 2012). The subsequent increase of organization allows addition of new elements, due to the increased flow capacity in the system, which again allows an increase of its quantity, and the cycle repeats. Those cycles are reflected in the data oscillations around the average fit values. In the case of CPUs, increasing α by making the transistors smaller and reducing the distance and friction between them allows more transistors to be added to the CPUs, increasing their quantity. On the other hand, adding more transistors allows more paths on the network, and miniaturization allows less energy and time to be used for one computation, increasing their quality. Both quality and quantity are in positive feedback, progressing in steps in an oscillatory mechanism. If any of the changes are outside of the stability limits, which are roughly the size of the observed deviations in the data, the system will be destabilized, and at even larger deviations, destroyed. Those deviations are necessary for α and Q to continue to grow, but their limits can be measured as the distance of those variations from the average fit of the data. We can label this average power law relation between α and Q as their "homeostatic" dependence. This means that the actual values of α and Q need to be close to those homeostatic values at each point, within the observed in the data limits of deviations around this dependence. Homeostasis implies a negative feedback between the actual values of α and Q and their homeostatic power law values given by the fit line, in order to prevent larger deviations from them, which can destabilize or destroy the system. The details of this oscillatory mechanism of growth and stability limits we will explore in consecutive papers.

Another reason for the observed oscillations can be that quality and quantity cannot change continuously. For example, the average unit action of a system will not change, if a half, or 99% of the length of a new flow channel is added, but only when it is completed. A structure cannot function until it is complete, which is in the definition of a complex system. Once the new channel is open, and the organization increases discontinuously, then quantity is allowed to accumulate, while the structure is the same – a cause for the oscillations in the data. As another example, enough elements need to group in a flow channel before they can overcome the minimization threshold for a certain large constraint, when the organization changes abruptly. Elements cannot be added to a channel that has a low action efficiency, because that will lead to jamming and the system will stop functioning. When the action efficiency is increased due to the constraint minimization, then new elements can flow through the system, without causing jamming, and which can further minimize larger constraints, and so on in a positive feedback. The addition of new elements to a system also happens discretely: the size of the atoms changes discretely, by adding new nucleons; the size of molecules, by adding new atoms; and the size of organisms by adding new cells; and the size of societies, by adding new members and structures. Due to this discreteness in the nature of the increase of the quality and quantity, the data do not increase monotonously but in steps.

Moreover, we also do not imply that the oscillatory dependence is valid only for processes of increase of $\alpha$ and $Q$, even though our data are collected only through the growth phase of the system that we have chosen. It should have the same validity for processes of decreased $\alpha$ and $Q$, because the same mechanism should act on the way down as on the way up in organization. If either $\alpha$ or $Q$ are decreased for any reason, they bring changes to decrease the value of the other, since neither the system has enough quantity to minimize the constraints anymore, which decreases its quality, or if its quality is decreased, that decreases its capacity to hold enough quantity.

**Conclusions**

We applied a positive feedback model based on a system of differential equations between the organization (quality, complexity) and the total amount of action in the system (quantity, size). We solved to obtain equations for exponential growth of both organization and total amount of action of the system, and a power law relation between the two. We used a sample system, which is the Core Processing Unit (CPU) of computers, to compare to our model and find a good agreement. This confirms the size-complexity rule (Bonner, 2004; Bell and Mooers, 1997; Carneiro, 1967) and efficiency-size observations (Bejan 2011, Kleiber-1932, West 1999) and leads to the conclusion that the decrease of unit action based on the principle of least action is connected to an increase of the total amount of action in a system. Each cannot continue its change without a change in the other within certain limits. Increasing the total action is necessary to do the work to minimize unit action, as less unit action is necessary to keep a larger system functioning as a whole. A system needs time and energy to self-organize, so smaller unit actions are achieved at later stages of its evolution and in larger systems. Decrease of action per unit motion is connected to the total amount of energy and time, i.e. total action in the system, because if the system does not have enough energy or time it cannot do the work to minimize the constraints. A system needs less unit action in order to transmit larger energy flows and to exist for a longer time. We can look at energy and time relationships separately, in order to understand them, but ultimately they are connected through action.

The principle of least action expanded for a complex system contributes to the explanation of the mechanism of increase of organization through quantity accumulation and constraint and curvature minimization with a final attractor state: the least average sum of actions of all elements and for all motions. The meaning of attractor in this case is a target state providing directionality for the process. Oscillations are important as they can explain that variation of one of the functions is necessary to cause a change in the other. There are possibly two principles ruling the self-organization of complex systems: the least unit action principle, and the most total action principle, both of which are inseparable. This mechanism is consistent with and potentially could explain other measurements in very different types of systems from physics, chemistry, biology, ecology, economics, cities, network theory and others, where the same power law dependence between qualitative and quantitative measures of the systems are observed. This understanding can help describe, measure, manage, design and predict future behavior of complex systems to achieve the highest rates of self-organization to improve their quality.

We recognize that there are limitations to the model and data. The mathematical model needs to be expanded, to include periodic oscillations of $\alpha$ and $Q$ around the homeostatic level observed in the data. It is plausible that the increase of total action drives a series of phase transitions decreasing the unit action, expressed as the oscillations of the data for $\alpha$ and $Q$ around the power law average, and vice versa. A model of series of logistic steps seems promising here, which have to be a solution of the underlying differential equations for positive and negative feedback loops. The data need to be expanded to a larger time interval to include the changes between different computing technologies for the last more than 100 year previously to CPUs' appearance. At the transitions between computing technologies, such as mechanical, electromechanical, lamp, transistor and integrated circuits, larger changes in action per computation may occur. With computing components decreasing by orders of magnitude in size and energy consumption, the unit action seem to decrease more abruptly at the transitions compared to between them, which remains to be investigated. There are indications for a possible super-exponential nature of $\alpha$ and $Q$ increase at longer time intervals. The exploration of increase of organization in complex systems is still in the early stages, and there is a lot to be discovered about its mechanisms. One interesting question is whether some of the constants may appear in other complex systems, which will make them universal, and increase the predictive power of the model. We will also explore the justification of the use of other, sometimes more easily accessible measures of quality and quantity or the dynamics of self-organization in general. Such are the amount of total flow of events in the system, the size and number of constituent elements, and others. We invite people who consider this topic interesting to join in this effort to decipher the code of progress in nature.

**Acknowledgements.**

The authors would like to thank Assumption College, for financial support and encouragement of this research.